\documentclass[english,twocolumn,APS]{revtex4}
\usepackage{amssymb}
\usepackage{dcolumn}
\usepackage{bm}
\usepackage{amsfonts}
\usepackage[T1]{fontenc}
\usepackage[latin1]{inputenc}
\usepackage{babel}

\begin{document}

\title{The Impact of the Higgs on Einstein's Gravity}

\author{M.D. Maia\\
University of Brasilia,  Institute of Physics, Brasilia 70910-900, maia@unb.br
\\\&\\
 Valdir  B. Bezerra\\ 
Federal University of Paraiba, Department of Physics,  Joao Pessoa, 58051-900, valdir@fisica.ufpb.br}

\date{\today}

\begin{abstract}
We present  an updated  review  of Kraichnan's  derivation of  Einstein's  equations   from quantum field  theory,  including the period after the  discovery of the Higgs  mechanism. Gravitation in the Einstein sense is  seen to be  renormalizable,
and  consistent with the Standard  Model of  Fundamental Interactions.
\end{abstract}

\maketitle

\section{The Gravitating Quantum Fields}

The derivation of  Einstein's  equations  from  the  spin-2  quantum field theory  was first obtained by  Robert  Kraichnan\cite{Kraichnan1955,Kraichnan1956},   an assistant of Einstein  at  Princeton  during the period  1949-1950. 
 According to the story\cite{Feynman1955}, Einstein suggested  to Kraichnan  an  investigation  on  the  emergence of  singularities from the  non-linearity of the  gravitational equations. 

The purpose of  this note is  to show  that   Kraichnan's  derivation of Einstein's equations implies that  Einstein's gravitational  theory  is  renormalizable,  offering a  definitive and unique solution   to the  long standing  search  for quantum gravity, including  its  implications to the   Standard  Model of  Fundamental Interactions (The Standard Model for short). We  start below  with a
brief  review  and update  of  Kraichnan's  derivation.  In  the next  section we show  why  that  result  implies in perturbative  quantum gravity. In section 3  we  suggest  a  way to  include  Einstein's gravitation in the  Standard Model,  using the Higgs mechanism.  

We begin when  Kraichnan's  noted   the  similarity between  the linear  gravitational  wave  equations and  the  wave  equation for  a massless  spin-2  quantum field  in the Minkowski space-time.
The  Lagrangian  of a  spin-2   field  $h_{\mu\nu}$  was  originally  derived by   Fierz and Pauli\cite{FierzPauli}  as (all Greek indices   $\mu, \nu...$  run  from $0$   to $3$. Unless explicitly indicated all small case Latin  indices   run from  1 to 3.)
\begin{equation}
\begin{array}{ll}
{\cal L}_{Fierz-Pauli}\! =\!\frac{1}{4} [
 h_{,\mu}h^{,\mu}\! -\!
h_{\nu\rho,\mu}h^{\nu\rho,\mu} -2h_{\mu\nu}{}^{,\mu}h^{,\nu}
+\vspace{1mm}\\
  2 h_{\nu\rho,\mu}h^{\mu\rho,\nu}- m^2 (h_{\mu\nu}h^{\mu\nu} - h^2)],  \label{eq:LagrangianFP}
\end{array}
\end{equation}
where   we have denoted $h=\eta_{\mu\nu}h^{\mu\nu}$  and the  field   potential energy is  $U=(h_{\mu\nu}h^{\mu\nu} - h^2)$. 
The observables of such field were defined as  being the non-zero spatial components  $h_{ij}$,  satisfying the  minimal energy condition  
\[
\frac{\partial U}{\partial h_{\mu\nu}}= h^{\mu\nu}-h\eta^{\mu\nu}=0.
\]
 With this condition,  it follows  that $h=0$  and  one  degree  of freedom is  eliminated. Consequently   the Fierz-Pauli  field  has  only  five non-zero independent  degrees of  freedom (dof). From  the  spin-statistics  theorem  we infer also that  $h_{\mu\nu}$  is  a  spin-2 field.  The Fierz-Pauli  field equations  are $(\Box^2-m^2)h_{\mu\nu}=0$  and in the  massless case they become  simply
\begin{equation}
\Box^2 h_{\mu\nu}=0. \label{eq:FierzPauli} 
\end{equation}

On the other hand,  the observables of Einstein's  gravitational  field are  given by the  real non-zero solutions 	 of the eigenvalue equations of the  curvature tensor\cite{Pirani}.	
\begin{equation}
R_{\mu\nu\rho\sigma}X^{\rho\sigma}= \lambda X_{\mu\nu}, \; \mbox{no sum on} \;\rho\; \mbox{and}\; \sigma, \label{eq:eigenvalues}
\end{equation}
where  $X^{\mu\nu}$  are   the  corresponding eigenvectors.
Since these eigenvectors are  anti-symmetric   2-forms,  there  are at  most  six  independent solutions, corresponding to  six  real  eigenvalues, including the trivial  solution  $\lambda =0$, which  corresponds to   the  vacuum state  of the gravitational  field.  Thus,  we are left also  with only five non-trivial eigenvalues $\lambda$  to be measured by  the  observers,  so that  Einstein's gravitation has $dof=5$   or  equivalently,  it is  a   spin-2 field. 
  
The  linear  gravitational  wave equations written  in the de Donder coordinate gauge are
\begin{equation}
\Box^2  \psi_{\mu\nu}=0,\;\;\;\;  \psi_{\mu\nu}=h_{\mu\nu}-\frac{1}{2}h \eta_{\mu\nu}.
\label{eq:lineargw}
\end{equation}
Comparing  with  Eq. (\ref{eq:FierzPauli}), we  find  that if   the  above  $h_{\mu\nu}$   coincide with  the  spin-2  field   defined by  Eq. (\ref{eq:FierzPauli}),  then  the  above   wave  equations  becomes  identical  to  the  Fierz-Pauli wave equations Eq.  (\ref{eq:FierzPauli})  in an arbitrary  reference frame.  
Such independence of  a  specific  coordinate gauge  suggests  that  gravitation  could  be  obtained from    sequence of small perturbations of the Minkowski metric by the  Fierz-Pauli   field $h_{\mu\nu}$ as
\[
g_{\mu\nu} =\eta_{\mu\nu} + \delta h_{\mu\nu} + \delta h^2_{\mu\nu} +\cdots.
\] 
After  a  laborious sequence of   calculations, the comparison of   each term  of  such  perturbation  with   an  arbitrary conserved   energy-momentum tensor  $T_{\mu\nu}$,   Kraichnan  obtained  Einstein's  equations,   with the only  difference that   the  coupling  constant   with   the  source  is an  arbitrary   proportionality  factor  $\kappa$,  not related  to the Newtonian gravitational constant $G$: 
\begin{equation}
R_{\mu\nu}-\frac{1}{2}R g_{\mu\nu}=\kappa T_{\mu\nu},
\label{eq:EK}
\end{equation}
Such  equation  is defined in the  Minkowski  space-time,  but from  this point on,  Kraichnan  just followed   Einstein's geometrization of  the gravitational  field:  Define  a new  space-time whose metric  $g_{\mu\nu}$  is a  solution of Eq. (\ref{eq:EK}) written  in  an  arbitrary  reference frame;   introduce  the  principles of general covariance and  of  equivalence; and  finally  define  the  observables  of  the  gravitational field (the  space-time  curvature tensor).  

Notice that  we could  also have  derived directly   the  Ricci  scalar using  Kraichnan's   perturbative  construction,   subsequently obtaining   the   Einstein-Hilbert Lagrangian  $R \sqrt{-g}$ and finally  applying  the  variational principle 
\begin{equation}
 \frac{\delta}{  \delta g_{\mu\nu}}\int{(R-\kappa{\cal L}_{source})}\sqrt{-g}dv =0.
 \label{eq:EH}
\end{equation}
Therefore  we may  infer that  Kraichnan's  result   expressed by Eq. (\ref{eq:EK})  is  unique  up to  the addition of a  divergence-free term. It  also   suggests  that  the  Einstein-Hilbert  Lagrangian  can be regarded  as a non-linear version of the  Fierz-Pauli variational principle  for  massless  spin-2 fields.   

Similar  derivations of  Einstein's  gravitation from  quantum  field theory were  also obtained by  several  authors in the period from  1950-1970, like  S.  Gupta, showing that  the Einstein tensor  could be  obtained from a  perturbation of  the Minkowski  metric  by a  quantum field\cite{Gupta1954}.  Richard Feynman in his  Lectures on Gravitation at Caltech also attempted  to derive   Einstein's  gravitation  from a  spin-2  quantum  fields,  starting from   Feynman's  diagrams.  A comprehensive analysis  of these  results  can be  found in the  foreword by  John Preskill and Kip Thorne to  Feynman's  Lecture Notes on Gravitation\cite{Feynman1955}.

Along a  similar  line,  it was  also  suggested that  a short range  form of  gravitation could  result from the  perturbations of the  Minkowski metric by a  massive spin-2 field\cite{vDam,Zakharov}, leading to what is  known today as massive gravity\cite{Claudia}.  Accordingly,  Einstein's  massless gravitational  field  would  result from  the  zero mass  limit  of  such  massive gravity.   However, it  has been  argued that  such  massive  gravity would be  plagued  with  ghosts,  which  would remain  even after  the zero mass limit\cite{DeserBoulware}. 
 A  short range gravitational  field  was  also  suggested  as a  spin-2  field independent of  the  space-time  metric, similar to the  electromagnetic field produced by a  local  current\cite{Salam}.   In spite  of   the ongoing   debate on the existence  short range gravity,    here we concentrate  in the most  relevant  implication of  Eq. (\ref{eq:EK}),  perturbative  quantum  gravity.

\section{Quantum Gravity}

The renormalization   of  gauge fields was proved  by  'tHooft \& Veltman\cite{Veltman},   but the  same  procedure   did   not  work for  the gravitational field of  General Relativity. This  can be understood  from   a  fundamental difference   between the two  theories:  In  gauge  theory,  the  coupling  constant   $\textbf{g}$  appears in the covariant derivative  $D_\mu= \partial_\mu  + \textbf{g} \; A_\mu $, where  $A_\mu$   is the gauge field, solution of   the  Yang-Mills field  equations  
\[
D^\mu F_{\mu\nu}=  4\pi J_{\nu}, \;\;  D^\mu F^*_{\mu\nu} =0,
\]
where $F_{\mu\nu} =[D_\mu,D_\nu]$ and where  $F^*$  is  the  dual of  $F$.
As we see,   the values of  $\textbf{g}$  can  be used as  a renormalization parameter  ``before''  it is  committed to the  dynamical  equations.

On  the   other  hand,  in  General  Relativity  the coupling constant has  a  specific  value  $\frac{8\pi G}{c^4} $  with  physical   dimensions,   which was   fixed at  level  of the  dynamical   equations    
 \[
R_{\mu\nu}-\frac{1}{2}R g_{\mu\nu}= \frac{8\pi G}{c^4} \; T_{\mu\nu}.
\]
It follows tat   the Einstein gravitational  field   in  GR cannot be  perturbatively   quantized because  the  coupling constant is   fixed  by  the  dynamical  equations,   a   conclusion reached  by  Goroff-Sagnotti\cite{Goroff},

However,   in the case of  the  Kraichnan  derivation, the source   $T_{\mu\nu}$  and the corresponding  coupling constant $\kappa$  were  not  specified and they 
 depend of the scale  of  observations,  from  the  range of  quantum field theory,   to the classical Newtonian level where the sources  are predominantly composed of   ordinary matter,  up to the  cosmological scale  where  gravitation may  also  also  couple  with  dark matter in the universe.

Following a  reasoning   similar to that of   Goroff-Sagnotti,  the  gravitational  field   defined  by  Eq. (\ref{eq:EK})  can be  written  in  a form that simulates  a  gauge  field where the  covariant derivative  is  modified to    $ \nabla_\mu = \partial_\mu  +  \sqrt{ \kappa} \Gamma_\mu $, for   $\kappa\neq 0$,  and simultaneously  writing the  variational principle as 
\begin{equation}
\frac{1}{\kappa} \frac{\delta }{\delta g_{\mu\nu}} \int{R\sqrt{-g} dv}=\frac{\delta }{\delta g_{\mu\nu}}   {\cal L}_{source}\sqrt{-g}dv,  
\end{equation}
Consequently,  the   gravitational  field  resulting from   Eq.  (\ref{eq:EK})
is  renormalizable  in the sense  of  Goroff-Sagnotti.  

Notice that  in the vacuum  case,  ($\kappa=0$) the    equations   Eq. (\ref{eq:EK}) correspond to  the  non-linear  version of the  Fierz-Pauli  field  without any source,  whose interpretation may  be  that of   a  non-linear  quantum spin-2  field.   One    interesting  example of  such  quantum field  is  given by  the spherically  symmetric vacuum  solution of   Eq.  (\ref{eq:EK}),   describing a   quantum   black-hole,  with   Schwarzschild radius  $r_{Sch}=\frac{2GM}{c^2}$,  where $G$ is  replaced by  $\kappa$  and    $M$ cannot be identified   with a Newtonian mass.  However  at  the level of  quantum field theory  $M$ can  be  identified with a  quantum of energy $M=E_0= \hbar  c/\lambda $,  so that  the  
 `` quantum Schwarzschild  black  hole'' has the   Schwarzschild  radius
\[
r_{Sch} = \frac{2 \kappa \hbar  }{ c\lambda}.
\]
In principle,  such  quantum  black  hole    may be experimentally   detected  
by  adapting the experimental  setup    previously proposed  to detect   black holes  from  extra  dimensions  at the tev  energy scale.  In  that   extra dimensional case,  the  existence of  tev  black holes   was justified   precisely because in  higher dimensions  the Newtonian gravitational constant  does not  depend  on  $G$  and  it  must  be replaced by a  higher dimensional  coupling constant\cite{MyersPerry,Dimopolous,ADD,Seong,Chamblin}.  The novelty  here  is in  the fact  that  such  quantum  black hole  is defined  in the  four  dimensional space-time.

\section{Gravitation  in  the Standard Model } 
  
The derivation of   Eq. (\ref{eq:EK}) from quantum field theory  implies that the gravitational hierarchy ceases to exist,  and quantum gravity exists  at tev.   Thus,  at least in  principle gravitation has the  right to be  explicitly present  
in the  Standard  Model.  

After the  experimental evidence of the Higgs,  the  Higgs  mechanism  explains how the  Higgs boson and the  spin-1 bosons  $Z^0$  and  $W^\pm$  acquire  their masses.   All  these masses  belong to the  spectrum of   eigenvalues of  the  second  order   Casimir  operator  $C^2 =P_\mu P^\mu$,  where  $P_\mu$  denote  the translation operators  of the Poincar\'e group.   
If so, then the  consistency   with the Lie  algebra  structure of  the  Standard  Model,   require that the  Poincar\'e Lie  algebra ${\cal P}_4$ should explicitly combine with   the gauge  field symmetries,  something like    
\[
{\cal P}_4 \times U(1) \times  SU(2) \times  SU(3).
\]  
However,  such   symmetry  combination  stumbles in the  classic problem  of   high energy  physics, the no-go  theorem for the  mixing  of the Poincar\'e  and gauge  symmetries\cite{ORaifeartaigh}.   In  short,  that   theorem tells that  the action of ${\cal P}_4$  over any operator of  $SU(n)$ is  idempotent, in the sense that
  the (squared)  mass difference of  two  particles $a$  and  $b$ belonging to the same  multiplet of  an $SU(n)$ internal symmetry vanishes: $\Delta m^2 =   m^2_a - m^2_b =0 $,  which is    a blatant   disagreement with the observations. Such  symmetry  mixing problem   has never been solved. At the best it was  averted  by  replacing  the  Poincar\'e  Lie  algebra   by  a  graded Lie  algebra\cite{Coleman}.  In the following,  we will see   why  the  solution  of such  problem  is  implicit in  the  structure of  the  Higgs potential. 

Consider  the well known  interaction  of  the Higgs   with the  electroweak field,  described by  the  Lagrangian (here for  simplicity  we take  $\textbf{g}=1$)
\begin{equation}
\begin{array}{ll}
 {\cal L}= -\frac{1}{4}tr \left[ F^{\mu\nu}F_{\mu\nu}+ (D^\mu \phi)^\dagger (D_\mu \phi)\right.-
\vspace{2mm}\\
\left. \mu^2 ( \phi^\dagger \phi )  +\lambda( \phi^\dagger \phi )^2 \right], \;  \lambda>0 \label{eq:Higgs},
\end{array}
\end{equation}
where $D_\mu= \partial_\mu +i  A_\mu$  and  $A_\mu$ is  the electroweak  field  potential. The Higgs is   a complex  column 
\begin{equation}  
\phi  =\left(
\begin{array}{cc}\phi_1\\
\phi_{2}
\end{array}
\right),
\;\;
 \phi^\dagger= \left(\phi^*_1,\phi^*_{2}\right). \label{eq:doublet}
\end{equation}
The  Lagrangian Eq.  (\ref{eq:Higgs})  is invariant under the  Poincar\'e  groups and   under  the  electroweak symmetry   $U(1)\times SU(2)$. 
The Higgs potential is   $U(\phi_1,\phi_2)$,  so  that the  minimum energy  condition for the Higgs  field is 
\[
\frac{\partial U}{\partial  \phi}=\mu^2 \phi^\dagger  + 2\lambda (\phi^\dagger\phi)\phi^\dagger=0, 
\]
whose  solution  $\phi= 0$   is  naturally excluded, as  it represents  an  unstable vacuum.   Since   $\lambda > 0$,   we  obtain also an  infinite  number of  non-trivial   real  and stable  vacuum  solutions by choosing   $-\mu^2  =m^2$, where $m^2$ is the  Higgs mass.  Those  stable solutions are  represented  by  the  points of a  circumference $|\phi|^2=\phi_1^2 + \phi_2^2= a^2=  m^2/2\lambda$  located  at  the bottom of the  Higgs  potential.

The  exclusion of  the  false vacuum  $\phi=0$  has the  mathematical implication   that the  Higgs potential is  a  genuine revolution surface, generated by  the  global  rotation  group  $SO(2)_{Higgs\,global}$,  through the axis  $U(\phi_1,\phi_{2})$, which does not touch the surface. As  a  revolution surface, the  Higgs potential can be represented  as  ``cassino roulette'', where   the bottom circumference  is  the Higgs vacuum.
In this case there  are two spontaneously broken  symmetries: The first one breaks  the global symmetry $SO(2)_{Higgs\,global}$,  associating the  Higgs  mass  $m^2=-\mu^2$, which  has the same value  for  all  points  of that circumference. This  is a   consequence of  the  the nilpotent action of the Poincar\'e translations over $SO(2)_{Higgs\,global}$, implying in  a zero mass splitting.  That is, for  any  two  points  $a$  and  $b$   of the vacuum  circumference  we  obtain the same Higgs  mass:   
\[
{(\Delta m}_{Higgs})^2=(m_{Higgs\, a})^2-(m_{Higgs\,b})^2 =0.
\]
Secondly, while the  global symmetry $SO(2)_{Higgs\,global}$  remains  broken, the  local gauge  symmetry  $U(1)\times SU(2)$ is also  broken,  without the interference of the   nilpotent   action of the  Poincar\'e translations. 

The  current  issue  on the  possible  instabilities of  the   Higgs vacuum   refers  to  the possible quantum  fluctuation of the  Higgs potential  caused by quantum gravity,  producing  lower energy stable solutions\cite{Gregory,Bednyakov}.
Such  quantum gravity  interference on the Higgs potential  must be  consistent  with  the  Einstein-Kraichnan quantum gravity and with  the standard model,  where
the Higgs potential   may have  lower  energy  stable   vacuum states  as  surfaces of  revolution, looking like a rotating  wavelet.    

\vspace{2mm}

In conclusion,  Kraichnan's  derivation of Einstein's  equations  is  correct and  unique, and it has some interesting  consequences to  the  understanding  of Einstein's  relativistic  gravitation. Not only because gravitation can be generated by   a  massless  spin-2    quantum  field,  but also  because  the  emergence of  quantum   black holes, ultimately  answers  Einstein's   question  concerning the origin of  singularities: These quantum black holes are  evidences  that  the  gravitational  singularities  emerge from quantum field theory  as shown by  Kraichnan.

	Finally, General  Relativity is  still  with us  as  a   low  energy limit of the Einstein-Kraichnan  Relativistic  gravitation. In such limit,  the  sources of  the gravitational  field  are  essentially   composed of  ordinary  matter,  with  $\kappa =\frac{8 \pi G}{c^4}$.  

\section*{Acknowledgments}

VBB thanks Conselho Nacional de Desenvolvimento Cient\'ifico e Tecnol\' ogico (CNPq) for partial financial support of the research project no 305835/2016-5.  MDM  thanks   prof. S. Deser  for a valuable comment on a previous version of this report.  Both authors  wish  to add a  posthumous thanks to prof. Leite Lopes.

\end{document}